\documentclass[12pt,twocolumn]{revtex4-1}

\usepackage{graphicx}
\usepackage{dcolumn}
\usepackage{bm}
\usepackage{color} 
\usepackage{graphicx}
\usepackage{natbib}
\usepackage{epstopdf}
\usepackage{amssymb}
\usepackage{psfrag}

\begin{document}


\title{How does turbulence spread in plane Couette flows?}

\author{M. Couliou}
\author{R. Monchaux}%
 \email{couliou@ensta.fr}
\affiliation{%
IMSIA, CNRS, UMR 8193, ENSTA-ParisTech, Universit\'e Paris Saclay, Palaiseau, FRANCE.
}
\date{\today}
\begin{abstract}
We investigate the growth in the spanwise direction of turbulent spots invading a laminar flow in a plane Couette flow. Direct Numerical Simulation is used to track the nucleation of streaks during the spot growth. Experiment and direct numerical simulation allow us to study the velocity of the spot fronts and of the vortices observed at the spots' edges.  All these results show that two mechanisms are involved when turbulent spots grow: a formerly proposed local growth occurring at the spot spanwise tips but also in comparable proportion a global growth induced by large-scale advection identified in the present work.
\end{abstract}
\maketitle
\section{\label{sec:intro}Introduction}
\begin{figure}
\begin{center}
\includegraphics[trim = 0mm 0mm 70mm 0mm, clip,height=.85\linewidth]{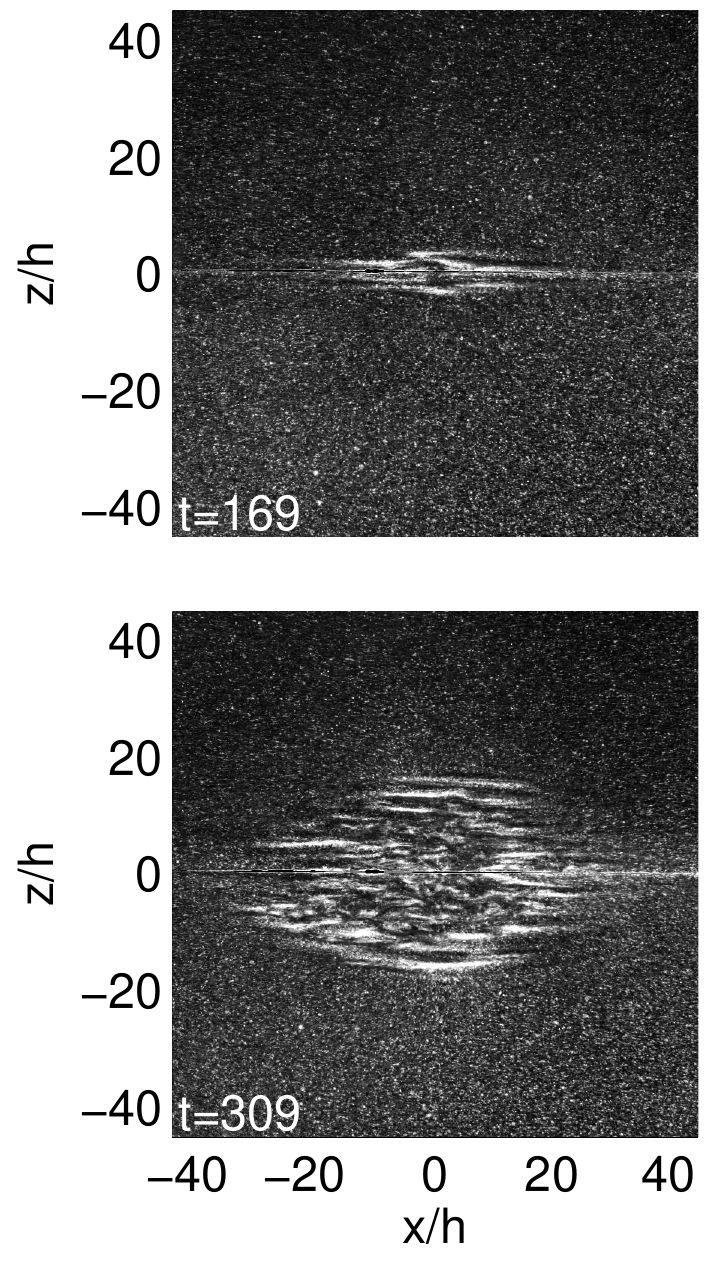}
\includegraphics[trim = 85mm 0mm 0mm 0mm, clip,height=.85\linewidth]{SpotSnapShots-eps-converted-to.pdf}
\end{center}
 \vspace{-.5cm}
\caption{Snapshots of a spot growing around a bead during a step experiment (sudden increase at $t=0$ of $Re$ from $0$ to $403$). Times are given in unit of $h/U$.}
\label{Snapshots_bead}
\end{figure}

\begin{figure}
 	\begin{center}
    \includegraphics[trim = 0mm 0mm 0mm 0mm, clip,width=.95\linewidth]{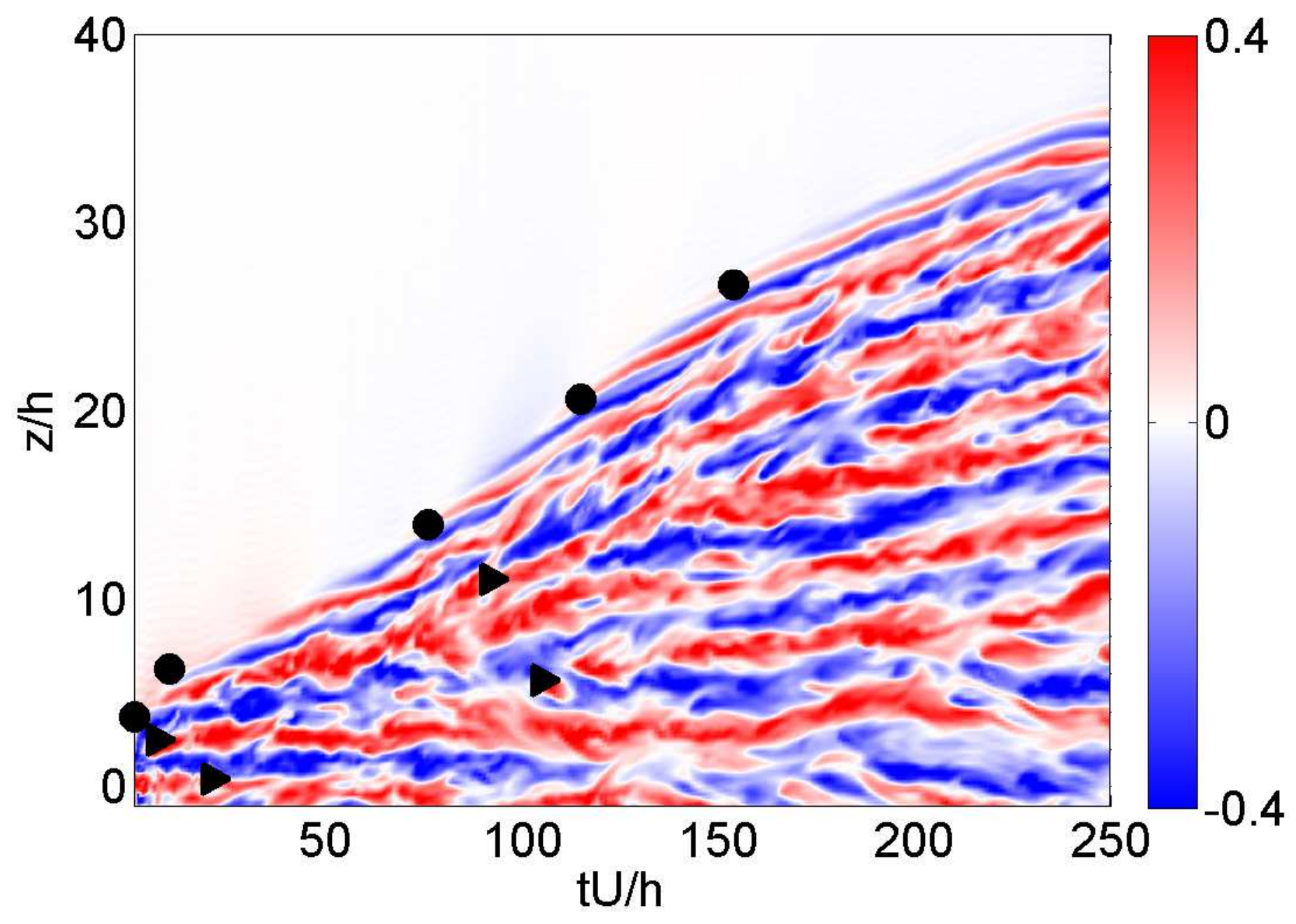}
	\end{center}
	 \vspace{-.5cm}
	\caption{Color online. $U_x$ space-time diagram for a DNS at $Re=380$. Nucleations of streak inside ($\triangleright$) and outside ($\circ$) of the spot are labelled. Only half the spot is presented in the spanwise direction.}
	\label{fig:DSTX_RE380_LX180}
\end{figure}

Transition to turbulence may happen through smooth supercritical scenarios involving successive symmetry breaking understood from linear stability analysis. Convection between differentially heated parallel plates or Taylor-Couette flow when the outer cylinder is at rest are typical examples of such transitions. Plane Couette flow, the flow between two parallel plates moving in opposite directions, experiences an abrupt transition to turbulence when the Reynolds number ($Re$) is increased that fails being predicted since all theoretical works conclude that the laminar base flow is linearly stable. This subcritical scenario involves the coexistence of laminar and turbulent phases for a range of Reynolds numbers lying between $Re_g$ below which any perturbation to the base flow asymptotically decays and $Re_t$ above which the flow is found to be homogeneously turbulent. This coexistence can take the form of organized oblique patterns observed experimentally and numerically in plane Couette flow \cite{prigent02_PRL,barkley05_PRL} but also in Taylor-Couette \cite{coles65_JFM,vanatta66_JFM} and in plane Poiseuille \cite{hashimoto09_THMT,tuckerman14_POF} geometries. So far, no clear explanation for the emergence of such patterns has been given. A way to understand the subcritical transition and the emergence of patterns is to focus on the growth of an initial turbulent germ in formerly laminar flow. When $Re>Re_g$, this germ will develop into a turbulent spot that will expand to eventually cover a significant fraction of the flow depending on the Reynolds number. See Fig. \ref{Snapshots_bead} for an illustrative picture showing the growth of a turbulent spot in plane Couette flow.\\
The fate of such a spot, and particularly its spreading rate, have been studied in plane Couette flow but also in boundary layers, e.g. in \cite{gadelhak81_JFM}, and in plane Poiseuille flow \cite{henningson94_JEM,lemoult13_JFM} to reveal the mechanisms responsible for its growth. In plane Couette flow, it is clear that above $Re_g$ a spot will grow at a rate increasing with $Re$ \cite{lundbladh91_JFM,tillmark95_EPL,dauchot95a_POF} and that large scale flows are present around it \cite{lundbladh91_JFM,schumacher01_PRE,lagha07_POF,duguet13_PRL}. Experimental works also report the existence of waves at the spanwise tips of the growing spot that would travel at a speed lower than the laminar-turbulent front \cite{tillmark92_JFM,daviaud92_PRL,dauchot95a_POF}. Nevertheless, these waves have never been studied precisely. Regarding the mechanisms involved in the growth, the local destabilization of the modified laminar flow at the spanwise tips of the spot has been suggested \cite{gadelhak81_JFM,henningson89_POF,dauchot95a_POF,tillmark95_EPL}. The corresponding growth rates have the correct order of magnitude but are still too small to fully explain the observed growth \cite{henningson87_JFM}. Large scale flows induced by the laminar-turbulent coexistence are also supposed to play a role \cite{schumacher01_PRE,lagha07_POF,duguet11_PRE,couliou15_POF}.
In wall-bounded shear flows at moderate $Re$, turbulence is composed of an alternation of high and low velocity streaks. In plane Couette flow, the typical streak width is $4-5h$, $h$ being the half-gap between the two walls. Consequently, a turbulent spot contains a given number of streaks and its growth corresponds to nucleations of new streaks. With this in mind, we plot in Fig. \ref{fig:DSTX_RE380_LX180} a space-time diagram obtained by extracting the streamwise velocity $U_x$ obtained from a DNS along a spanwise line centered on a growing spot in the midplane between the two walls. The streaks are clearly recognizable and we observe nucleations of new streaks at two different locations: at the laminar/turbulent boundaries of the spot ("outer streaks" indicated with '$\circ$', compatible with the local destabilization) but also inside the spot ("inner streaks" indicated with '$\triangleright$'), within turbulence itself. These two locations suggest two different underlying mechanisms. The purpose of our study is to demonstrate and discuss these mechanisms allowing the growth of a turbulent spot. We first describe experimental and numerical methods. We then present the measurements of the spot growth rate and of the velocity of the vortices that we observe at the spot edges. We finally discuss the implications of these results on the spot growth process. Confronting them to the quantification of inside/outside nucleations, we suggest a growth scenario governed by two mechanisms: a local and a global one.\\
\section{methods}
We define $x$, $y$ and $z$ as the streamwise, wall-normal and spanwise directions and we note $U_x$, $U_y$ and $U_z$ the corresponding fluid velocities. The plane $y=0$ corresponds to the plane at half distance between the two walls.
\subsection{Experimental set-up}
An endless plastic belt linking two cylinders enable us to approach the ideal plane Couette flow. One of the cylinders is connected to a servo-motor driving the system. The gap between the two walls equals $2$~$h=7.5$ mm and is adjusted by four smaller cylinders. This set-up is immersed in a tank filled with water. Plane Couette flow is achieved within a $L_x=800$~mm (along $x$) $\times$ $L_z=400$~mm (along $z$) area. The Reynolds number is defined in our experiment as $Re=Uh/\nu$, where $U$ is the wall velocity and $\nu$ the water kinematic viscosity. Temperature is monitored at an accuracy of $0.1$~$^{\circ }$. We estimated the thresholds $Re_g = 300$ and $Re_t$ $\simeq 410$. Flow visualizations are performed by lighting Iriodin particles with a laser diode combined with a polygon mirror rotating at high frequency to produce a laser sheet which illuminates the flow along a plane at a given $y$. Iriodin particles are micron-sized platelets which tend to align with the flow stream. These platelets respond to local flow changes and reveal laminar and turbulent regions: the turbulent state is indicated by quick fluctuations in the reflected light intensity while the laminar state corresponds to more uniform light reflections. Images at a resolution of about $2200\times 1100$ pixels$^2$ corresponding to an observation window of $800\times 400$~mm$^2$ --exactly fitting the area where the laminar Couette profile is achieved-- are captured by a PCO sCMOS Camera. Fig. \ref{Snapshots_bead} illustrates typical visualizations with both laminar and turbulent states coexisting. As in Bottin et  \textit{al.} \cite{bottin97_PRL}, the flow is permanently and locally perturbed by a small bead (whose diameter is $0.8$~$h$) placed in the $y=0$ plane and held by a $0.1$~$h$ diameter wire parallel to the streamwise direction to avoid any perturbation induced by its wake. According to \cite{bottin97_PRL}, the bead generates four pairs of streamwise vortices that, in our case, trigger reproducible growing spots. The bead has no noticeable influence regarding critical Reynolds numbers and transition time to turbulence. To observe the growth of a turbulent spot, we performed step experiments consisting in a sudden increase of $Re$ from  $Re=0$ to $Re=Re_{f}$. Experimental and perturbation devices and their validations are detailed in \cite{couliou15_POF}. In the next section Particle Image Velocimetry (PIV) measurements are used as illustration of large scale flows. They have been obtained from a standard Dantec PIV system that is also fully described in \cite{couliou15_POF}.
\subsection{Numerical simulations}
DNS of the Navier Stokes equations in a plane Couette flow are computed using the Channelflow software (\verb|http://channelflow.org|) written by John F. Gibson \cite{channelflow,GibsonHalcrowCvitanovicJFM08}. The code uses pseudo-spectral methods for the spacial discretization with a Fourier decomposition in the ($x$,$z$) directions and Chebyshev polynomials in $y$ direction. The boundary conditions are periodic in the ($x$,$z$) directions and no slip conditions are imposed  at the walls $i.e$ at $y=\pm1$. Numerical simulations are performed in a domain of size ($L_x=180$~$h$, $L_y=2$~$h$, $L_z=80$~$h$). In order to resolve all the relevant modes of turbulent flow in the $Re$ range studied, the numerical resolution is (768,33,384) dealiased modes in the ($x$,$y$,$z$) directions. A time-step of $0.01$ was used resulting in a CFL number less than $0.6$. A four pairs of counter-rotating vortices as the one used by Lundbladh and Johansson \cite{lundbladh91_JFM} is introduced as an initial disturbance to trigger turbulent spots. 
\section{Results}
\subsection{Nucleations of streaks:}
In wall-bounded flows, transitional turbulent areas are composed of streaks. Since the width of streaks is narrowly distributed around a given size $\lambda_c$, when a turbulent spot grows $i.e.$ when the size of the turbulent area increases, the number of streaks has to increase too. We performed several DNS of a turbulent growing spot in order to track the nucleation of new streaks. As a streak is an elongated structure in $x$, we decided to focus on the nucleation events along the centered spanwise line ($x$,$y$) $=$ ($0$,$0$). We use $U_x$ space-time diagrams along this line like the one shown in Fig. \ref{fig:DSTX_RE380_LX180}. A sliding average is applied over $5$ time steps and over $3$ spanwise coordinates. From the space-time diagram, positive ({\it resp.} negative) streaks are identified as regions where $U_x>0.04$ ({\it resp.} $U_x<-0.04$). The central line of each streak are tracked in time using a particle tracking algorithm. Trajectories lasting less than $10$~$h/U$ are discarded. As already mentioned, nucleations are found inside and outside the turbulent area as illustrated in Fig. \ref{fig:DSTX_RE380_LX180}. The typical number of nucleations during the growth of a spot for $Re$ between $340$ and $420$ is fluctuating between $17$ and $23$.
\subsection{Spatiotemporal dynamics}
\begin{figure}
\begin{center}
\includegraphics[trim = 0 0 0 0, clip,width=.98\linewidth]{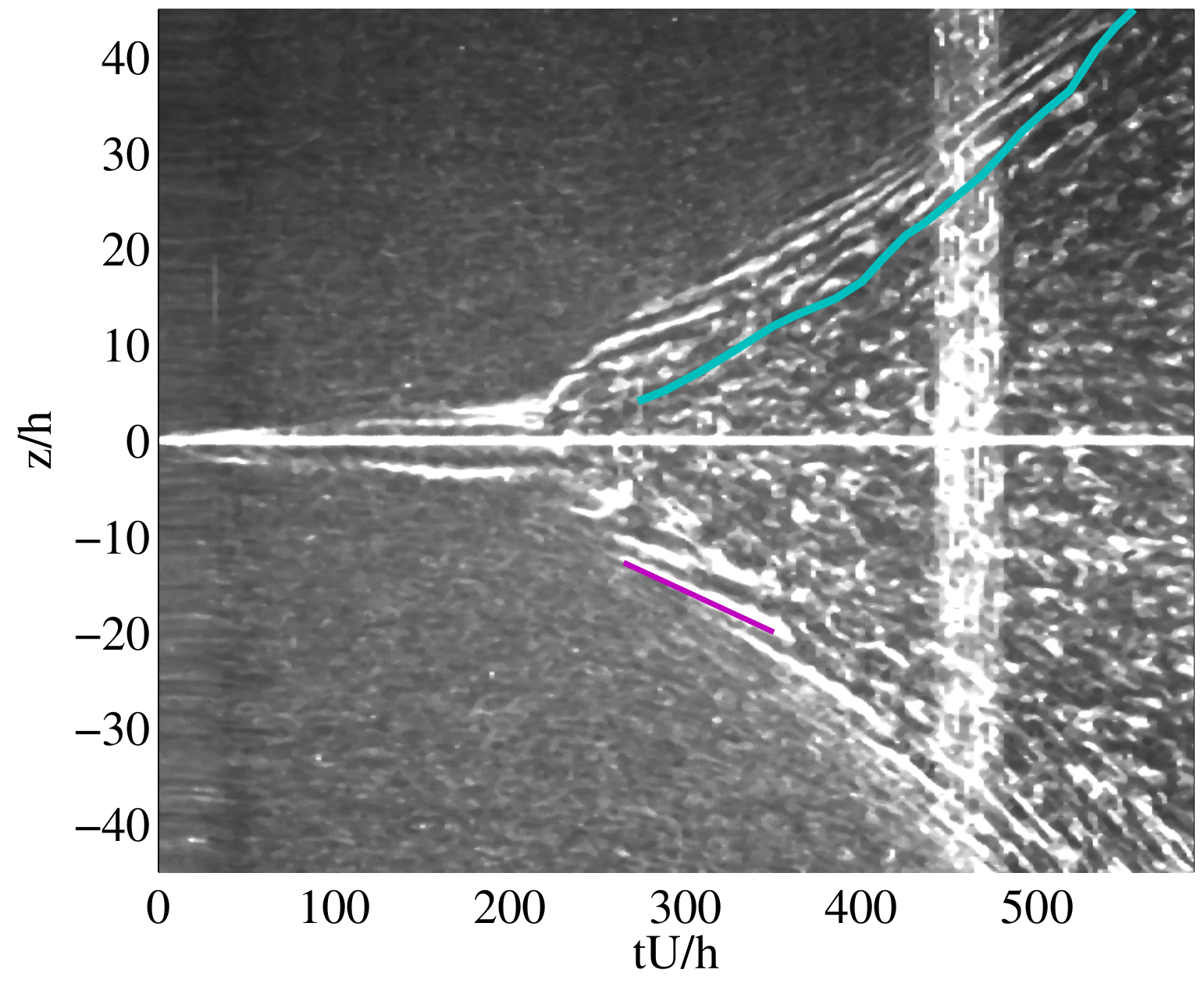}
\end{center}
 \vspace{-.5cm}
\caption{Color online. Space-time diagram from the experimental visualization corresponding to the growing spot presented on Fig. \ref{Snapshots_bead}. One of the fronts around the turbulent spot is indicated as a blue line and one vortex trajectory is marked with a purple line. Final $Re=403$ is reached after $10$~$h/U$ and, according to \cite{tillmark92_JFM}, the linear Couette profile is established after about $0.25 Re h/U$ corresponding to $100$~$h/U$ in this case. The white vertical band from $450$~$h/U$ to $480$~$h/U$ corresponds to the tape which closes the plastic belt.}
\label{fig:dst}
\end{figure}

\begin{figure}
\begin{center}
\begin{minipage}[h]{.55\linewidth}
\includegraphics[trim = 0mm 0mm 0mm 0mm, clip,width=\linewidth]{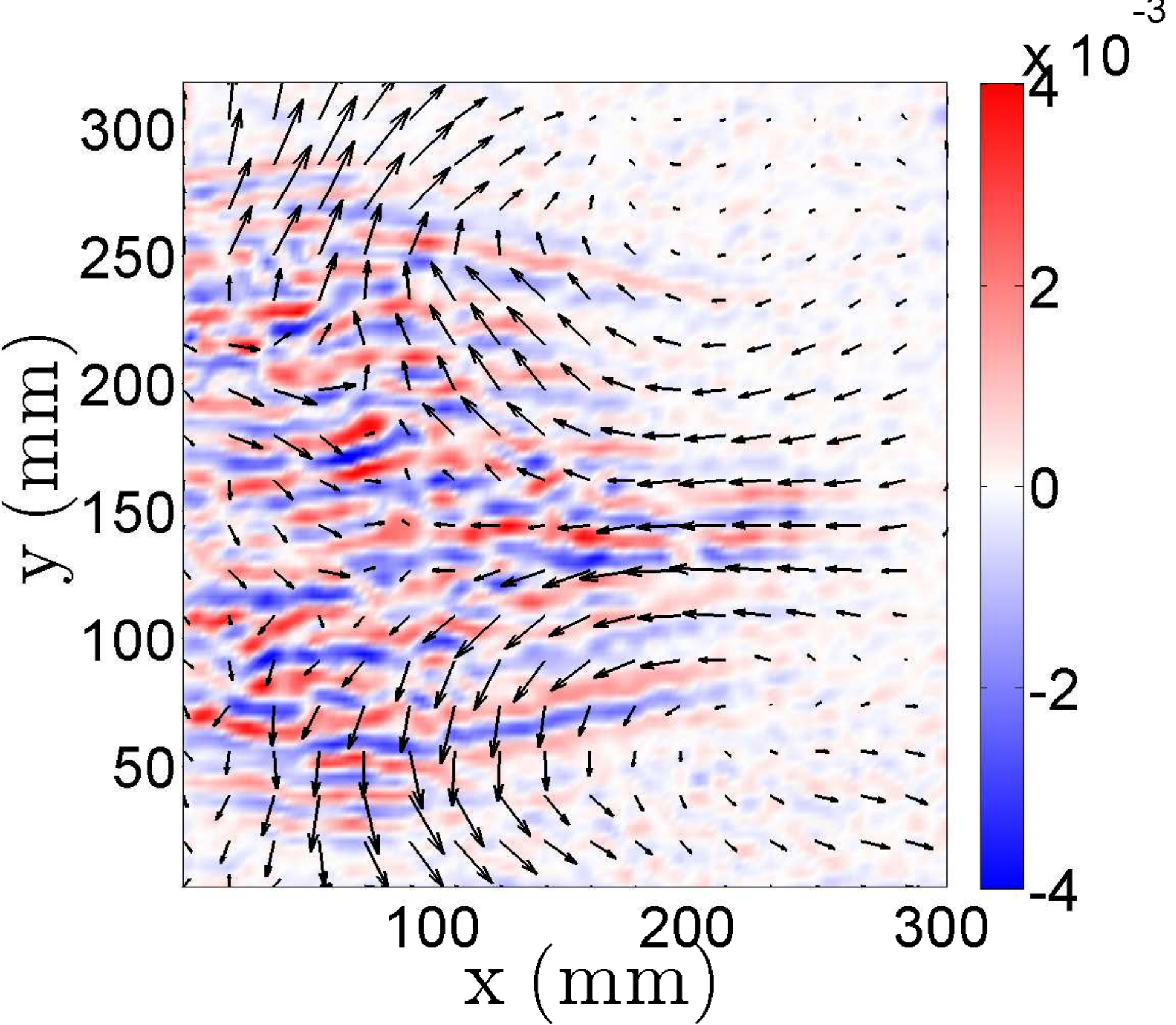}
\end{minipage}
\begin{minipage}[h]{.43\linewidth}
\includegraphics[trim = 0mm 0mm 0mm 0mm, clip,width=\linewidth]{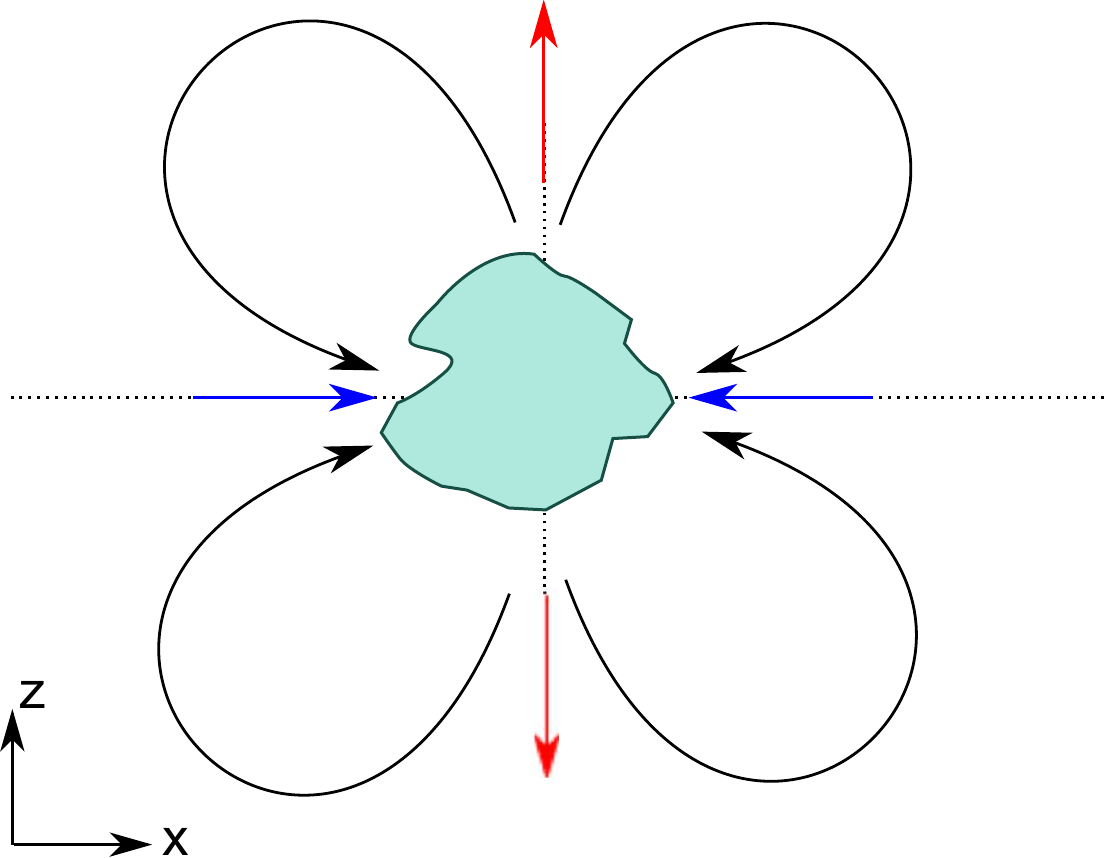}
\end{minipage}
\end{center}
 \vspace{-.5cm}
\caption{Color online. Left: contour of small scale wall-normal vorticity $\omega_y$ in a $(\vec{x},\vec{z})$ plane situated around $y=0$, large scale in plane velocity field is superimposed as arrows. Data are obtained from PIV measurements whose acquisition, post-processing and filtering are described in \cite{couliou15_POF}. Right: schematic representation of the turbulent spot (green surface) and of the corresponding large scale flow (arrows).}
\label{fig:LSF}
\end{figure}

Fig. \ref{fig:dst} shows the spatio-temporal evolution of a growing spot. The diagram is obtained from experimental visualizations by extracting the spanwise line crossing the bead. The time $t=0$ corresponds to the time when the step from $Re=0$ to $Re=403$ is performed. A first stage between $t=0$ and $t=210$~$h/U$ corresponds to a period when only a very local part of the flow is perturbed. Subsequently, a second stage is observed when ripples characterizing the turbulent state appear around the bead. From $210$~$h/U$ onward, the turbulent area invades the whole spanwise direction. Throughout the growth, long inclined coherent structures surround the turbulent spot and are visible between the laminar and turbulent phases. Several authors in former similar studies \cite{daviaud92_PRL,tillmark92_JFM} referred to them as waves. 
At these locations, it has been shown numerically and experimentally \cite{lundbladh91_JFM,lagha07_POF,duguet13_PRL,couliou15_POF} that the large scale flows induced by the laminar-turbulent coexistence are mostly oriented along the $z$ direction and are flowing away from the turbulent spot as sketch in Fig. \ref{fig:LSF}-b. This is further illustrated in Fig. \ref{fig:LSF}-a that presents large scale flow topology around a growing turbulent spot obtained from PIV measurement in our experiments (see caption for details). The large scale flow topology is thus exactly the one required to explain the "waves" as the simple advection of coherent structures located at the spot spanwise tips as already mentioned in \cite{duguet13_PRL}. From our numerical simulations, we have identified these structures as streamwise vortices that are actually also visible at the spanwise tips of the turbulent growing spot represented in Fig. \ref{Snapshots_bead} consistently with observations by Hegseth \cite{hegseth96_PRE}. To these vortices are associated streaks that can actually be seen in the numerical space-time diagram of Fig. \ref{fig:DSTX_RE380_LX180} as the outer streaks. At nucleation, they are smooth and straight before being subjected to a streak instability \cite{waleffe97_POF} turning these transient structures into turbulent streaks.
\begin{figure}[t!]
 \begin{center}
 	\includegraphics[height=.22\textwidth]{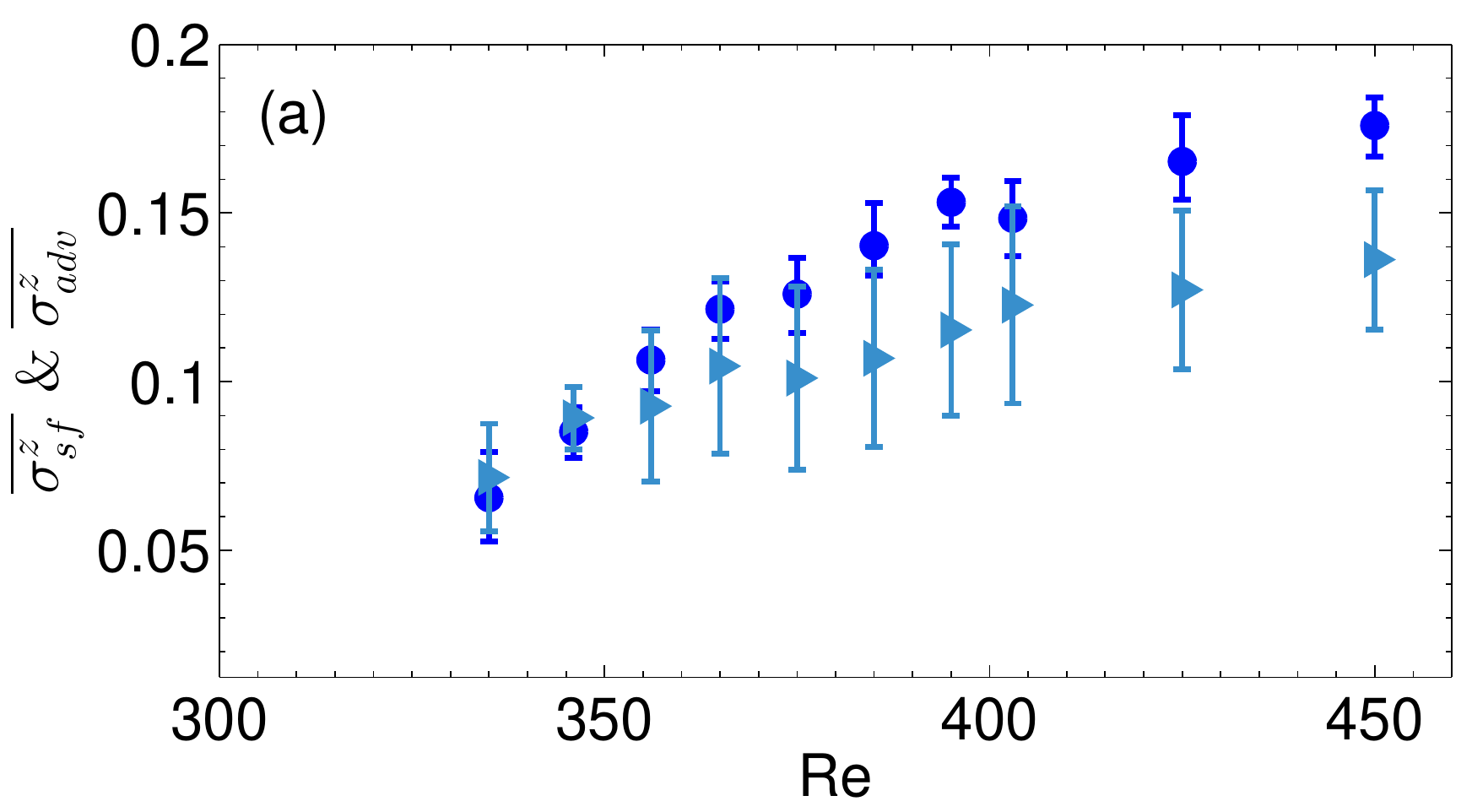}
 	\includegraphics[clip,trim = 0 0 0 0,height=.22\textwidth]{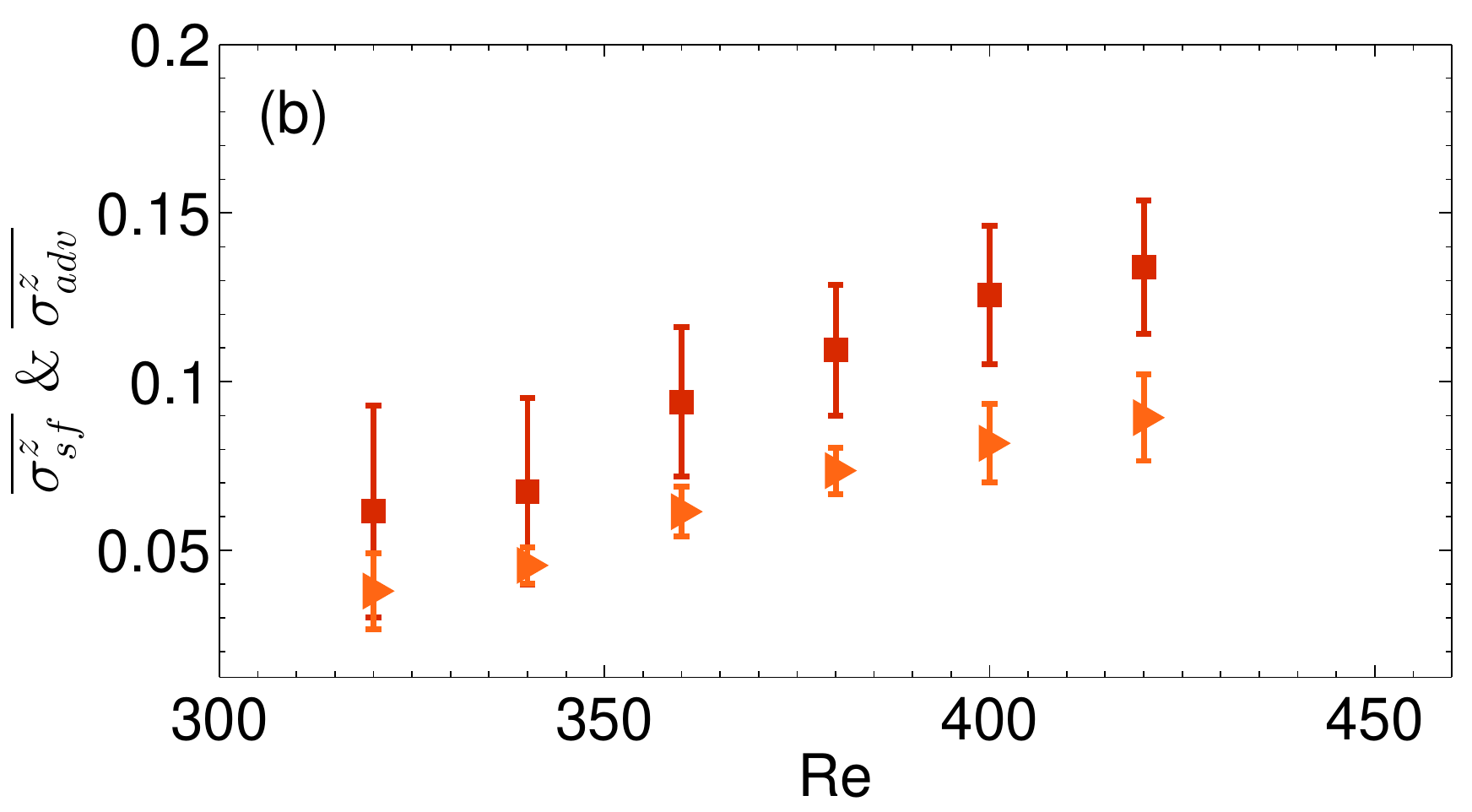}
  \end{center}
 \vspace{-.5cm}
\caption{Color online. Averaged velocity of the spot fronts $\overline{\sigma_{sf}^z}$ and of the edge vortex $\overline{\sigma_{adv}^z}$ as a function of $Re$. (a) $\overline{\sigma_{sf}^z}$ ($\circ$) and $\overline{\sigma_{adv}^z}$ ($\triangleright$) in the experiments. (b) $\overline{\sigma_{sf}^z}$ ($\square$) and $\overline{\sigma_{adv}^z}$ ($\triangleright$) in the simulations.}
\label{fig:edge vortex}
\end{figure}
\begin{figure}[t!]
 \begin{center}
 	\includegraphics[clip,trim = 0 0 0 0,height=.22\textwidth]{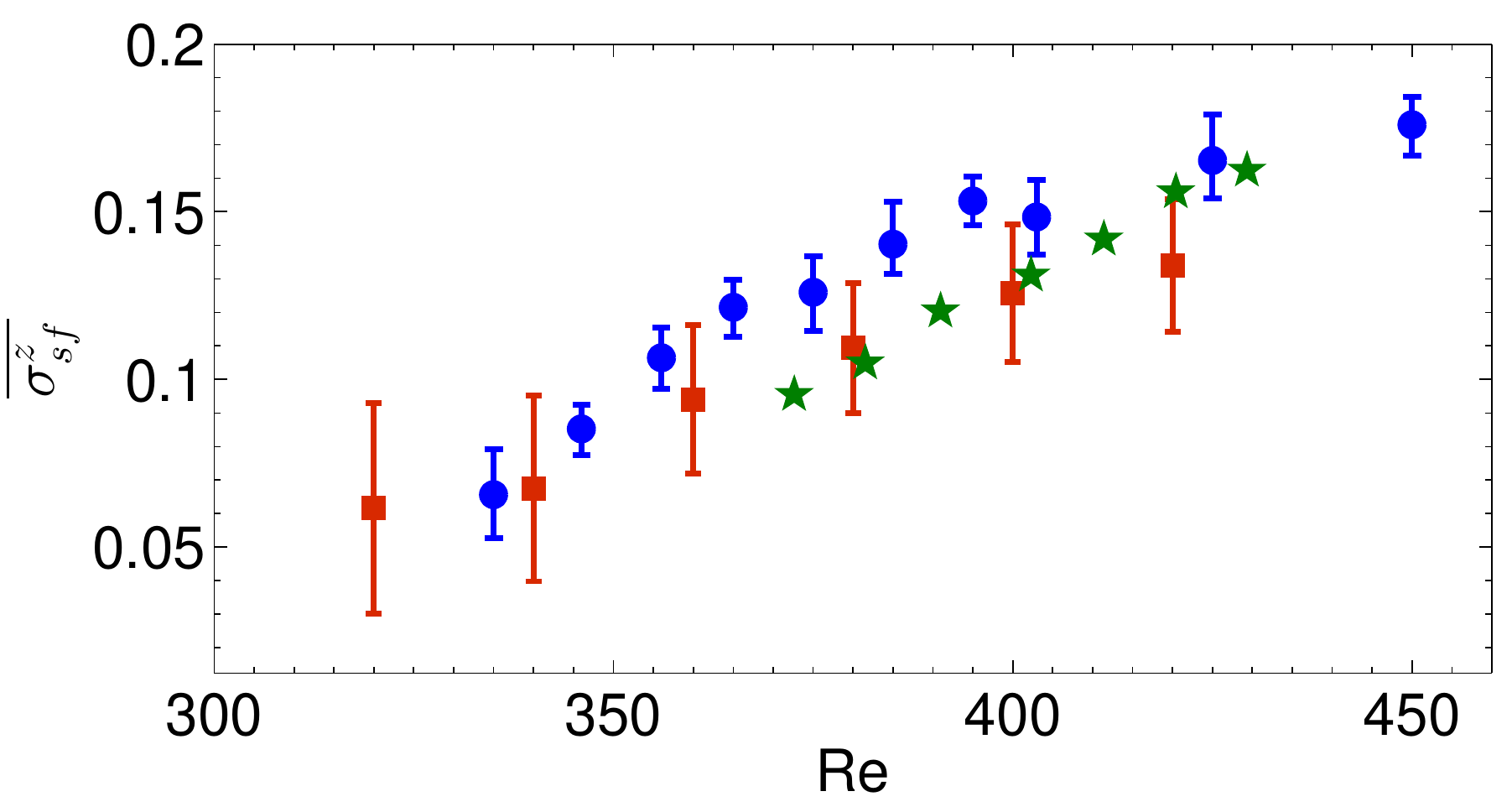}
  \end{center}
 \vspace{-.5cm}
\caption{Color online. Comparison between experiments, simulations and former work: $\overline{\sigma_{sf}^z}$ from the experiments ($\circ$), the simulations ($\square$) and from data by Dauchot and Daviaud \cite{dauchot95a_POF} ($\star$).}
\label{fig:saclay}
\end{figure}
\subsection{Front and edge vortex velocities}
We first focus on spot fronts and their velocities. From figure as Fig. \ref{fig:dst}, spot fronts, \textit{i.e.} the boundaries between the turbulent area and the edge vortices region can be easily identified. Their position have been manually tracked for each experiment and gathered when performed at the same $Re$. A typical track is shown in Fig. \ref{fig:dst} as a dark blue line. From the numerical data, the front is detected by a simple threshold on the velocity norm. The so determined front position hardly depends on the chosen threshold. Both in the numerical and experimental cases, the dispersion between different realizations is small enough to define an average front. This average front position at a given $Re$ is obtained by cubic spline interpolations with smoothing parameter $\lambda$. This cubic spline is then differentiated to obtain the front velocity as a function of time. Several values of $\lambda$ in a range between $\left[0.5;1\right]$ lead to slightly different velocity signals. These different signals are then time and ensemble-averaged to obtain the front averaged velocity at a given $Re$. The corresponding standard deviation is used as an error bar. To avoid finite size effects, we stop recording the front position when the spot fronts reach $z=\pm 36$~$h$. Regarding the edge vortices visible at the spot edge in Fig. \ref{fig:dst}, their positions are also manually extracted from the space-time diagram where they are considered as straight lines whose slope corresponds to their velocity. For a given $Re$, the edge vortex velocities estimated from all realizations are time and ensemble averaged. Fig. \ref{fig:edge vortex} gathers the evolution of front and edge vortex averaged velocities vs. $Re$ for both experiments (Fig. \ref{fig:edge vortex}(a)) and numercial simulations (Fig. \ref{fig:edge vortex}(b)). While both increase with $Re$, the edge vortex velocity $\sigma_{adv}$ is two third of the spot front velocity $\sigma_{sf}$ when $Re>350$. Fig. \ref{fig:saclay} presents the comparison of the total spot growth rate $\sigma_{sf}^z$ from our experiments and simulations to the results formerly obtained by Dauchot and Daviaud \cite{dauchot95a_POF}. The order of magnitude as well as the dependency on $Re$ are very similar between the three cases even if our experimental data show slightly faster fronts. It has to be noted that our experiment is significantly wider in the spanwise direction than the one used in \cite{dauchot95a_POF} ($400$\~mm \textit{vs.} $250$\~mm). A possible explanation will be given in the concluding remarks.
\subsection{Two growth mechanisms}
\begin{figure}
 \begin{center}
 \includegraphics[width=.4\textwidth]{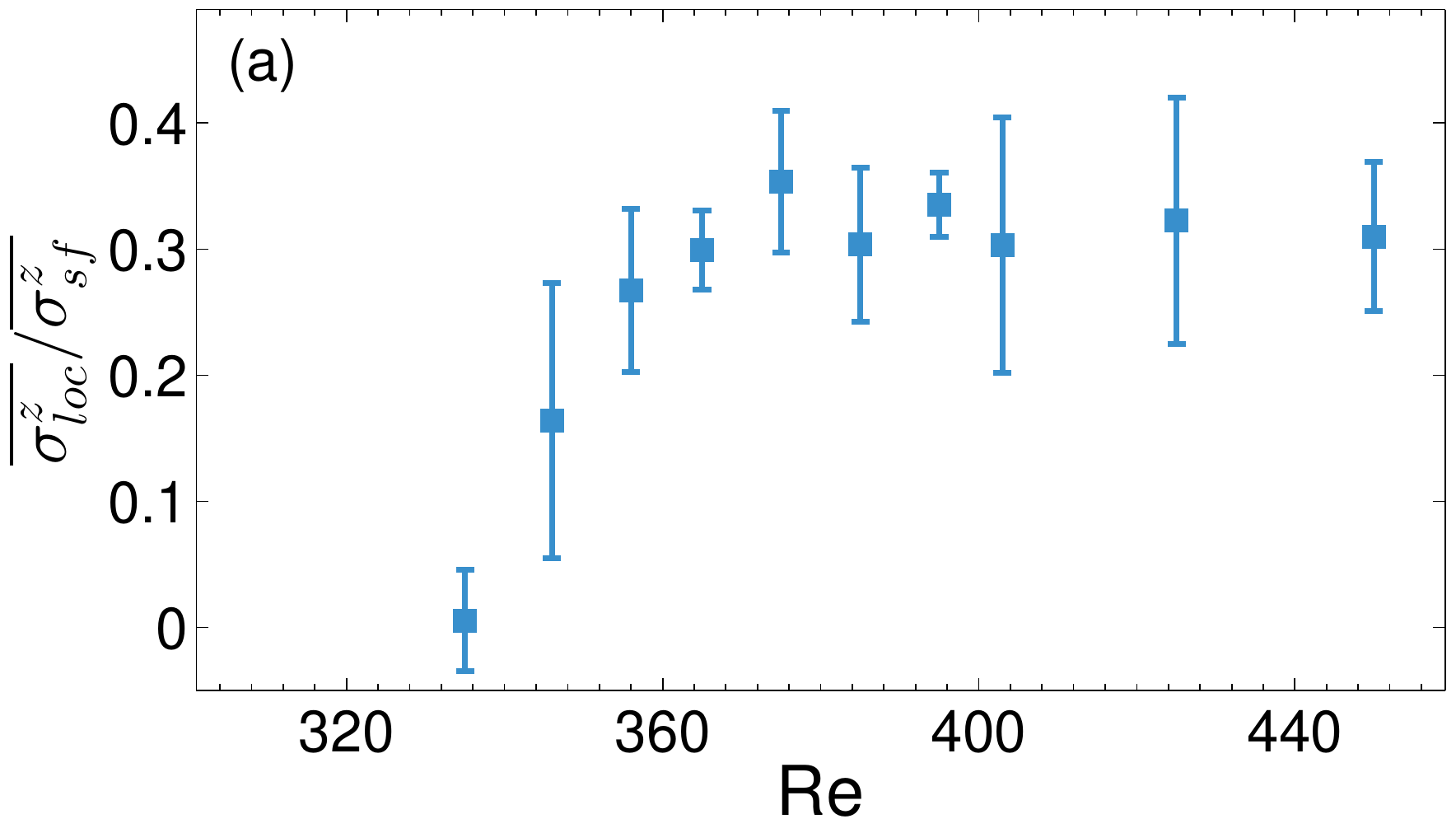}
 \includegraphics[width=.4\textwidth]{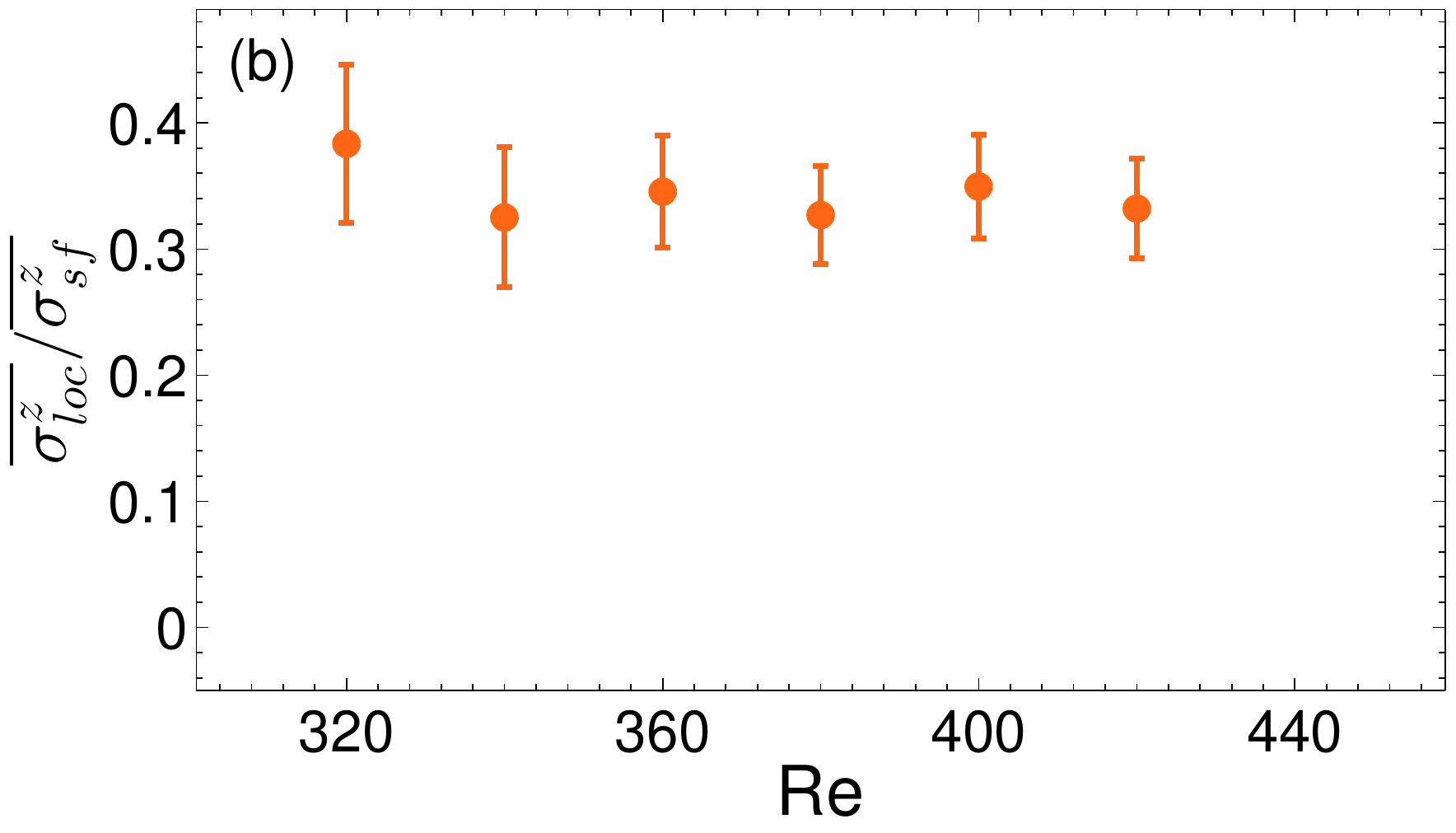}
 \includegraphics[width=.39\textwidth]{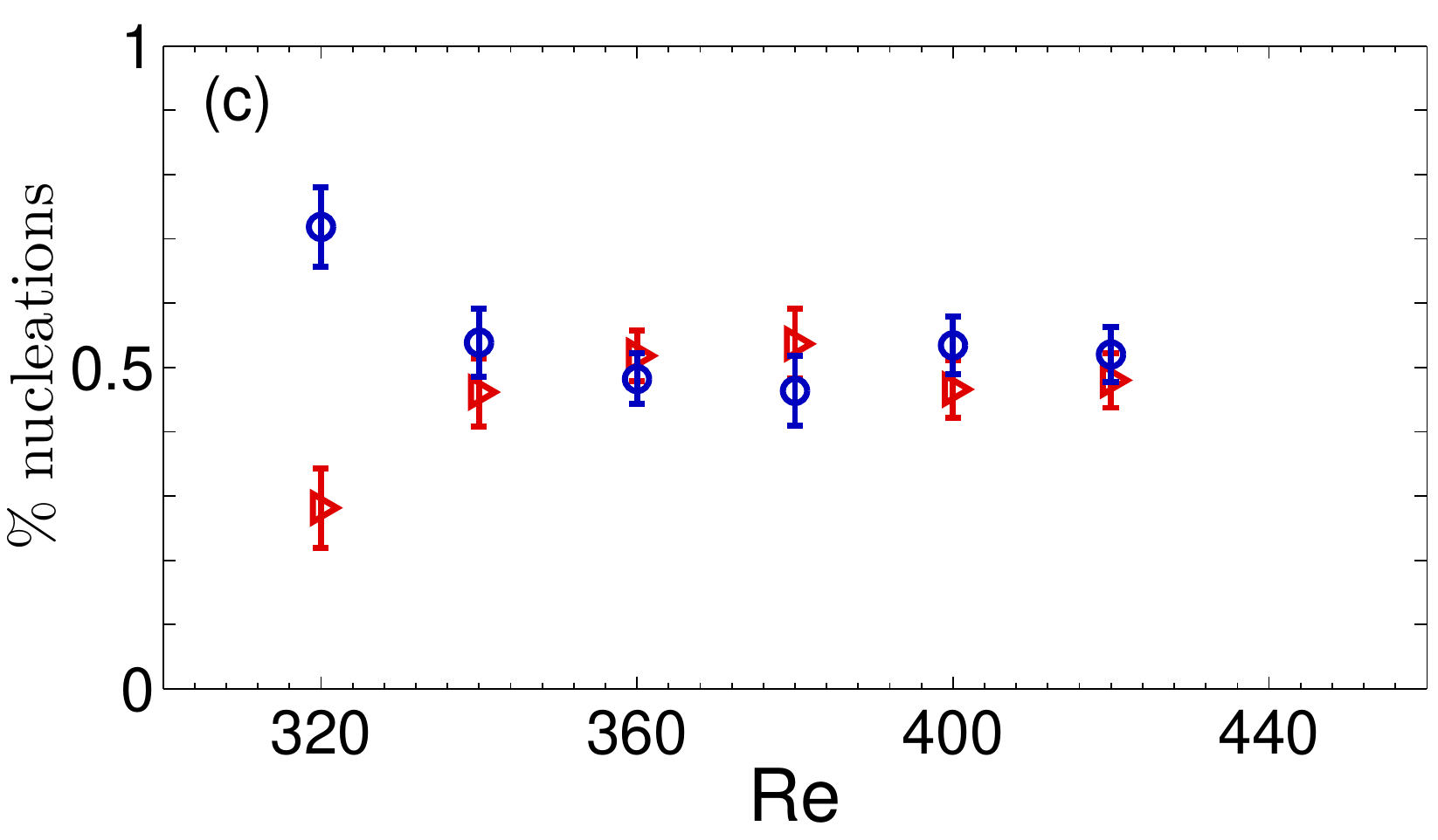}
 \end{center}
 \vspace{-.5cm}
\caption{Color online. Ratio between $\overline{\sigma_{loc}^z}=\overline{\sigma_{sf}^z}-\overline{\sigma_{adv}^z}$ and $\overline{\sigma_{sf}^z}$ as a function of $Re$ in the experiment (a) and in the DNS (b). (c) Fraction of nucleation of new streaks detected outside ($\circ$) or inside ($\triangleright$) the turbulent spot as a function of $Re$ in the DNS.}
\label{fig:mecha}
\end{figure}

The difference between $\sigma_{adv}$ and $\sigma_{sf}$ can be linked to two different growth mechanisms. The usually evoked local growth at the spot spanwise tips corresponds to new streaks nucleated {\it outside} the turbulent phase at a rate $\sigma_{loc}$ and to a growth without advection at work as in the narrow domains studied in \cite{duguet11_PRE}. The edge vortices advection suggest an additional growth mechanism in the spanwise direction. At the spot spanwise tips, the large-scale flows induced by the laminar-turbulent coexistence, are oriented spanwisely and outward the spot. Due to this topology, they stretch the spot in the spanwise direction by advecting it. The edge vortices are tracers of this advection and allow us to measure the corresponding growth rate as $\sigma_{adv}$. If the spot spreads without any nucleation of new streak, the local wavelength of its constitutive streaks increases slightly. New streaks have to be nucleated {\it within} the turbulent phase to restore the favored state consisting of streaks distant of almost $\lambda_c$. This second mechanism is illustrated in Fig. \ref{fig:DSTX_RE380_LX180}.\\
We have identified two growth mechanisms a local and a global one whose rates are $\sigma_{loc}$ and $\sigma_{adv}$ respectively. The total spot spreading rate can thus be written as $\sigma_{sf}=\sigma_{loc}+\sigma_{adv}$. From the measurements of $\sigma_{adv}$ and $\sigma_{sf}$ presented in Fig. \ref{fig:edge vortex} we have an estimation of $\sigma_{loc}$ as $\sigma_{sf}-\sigma_{adv}$. To compare the relative importance of both mechanisms, the relevant quantity appears to be $\sigma_{loc}/\sigma_{sf}$ whose evolution with $Re$ is shown in Fig. \ref{fig:mecha}. An increase of  $\sigma_{loc}/\sigma_{sf}$ at low $Re$ followed by a saturation around $0.35$ at higher $Re$ is observable. This has to be compared to the ratio of nucleations occurring $inside$ or $outside$ the turbulent spots as obtained numerically from space-time diagrams as the one shown in Fig. \ref{fig:DSTX_RE380_LX180}. This ratio also presented in Fig. \ref{fig:mecha} is almost constant and equal to $0.5$ when $Re>340$, revealing that both mechanisms contribute equally to the spot growth.
\section{Conclusion}
\paragraph*{}We investigated the spread of turbulent spots experimentally and numerically. We have measured spot fronts velocity as $\sigma_{sf}^z$ and found values comparable even if slightly larger in the experimental case to the one measured by Dauchot and Daviaud \cite{dauchot95a_POF}. Transient vortices are observed at the laminar-turbulent boundaries. We have tracked them at the spot spanwise tips from experimental visualizations and numerical simulations. Their nucleation is associated to a local growth mechanism while their drift is interpreted as a global advection by large scale flows. The total spot spreading rate $\sigma_{sf}$ therefore results from two contributions: on the one hand $\sigma_{adv}$ which indicates the contribution of advection and, on the other hand $\sigma_{loc}=\sigma_{sf}-\sigma_{adv}$ which corresponds to the local growth mechanism occuring at the spot spanwise tips. They are both $Re$ dependent. The latter corresponds to nucleations of new streaks $outside$ the turbulent spot while the former promotes nucleations of new streaks $within$ the turbulent phase as observed on the space-time diagram presented on Fig. \ref{fig:DSTX_RE380_LX180}. The difference between the values of $\sigma_{sf}^z$ measured here and the one measured in \cite{dauchot95a_POF} may arise from the fact that our experiment is significantly wider than theirs. As a result, large scale flows have more space to develop in our setup and thus their contribution can be significantly more important. Unfortunately $\sigma_{adv}^z$ has not been measured in \cite{dauchot95a_POF} which does not allow us to test this hypothesis. The mechanisms responsible for the nucleations of new streaks at the spot spanwise tips are out of the scope of the present article. They may nevertheless arise through an instability mechanism that could be the one proposed by Hegseth \cite{hegseth96_PRE} or the growth by destabilization suggested by several authors in wall shear flows\cite{gadelhak81_JFM,henningson89_POF,dauchot95a_POF,tillmark95_EPL}. This local destabilization could result from the local modification of the laminar flow by the large scale flows at the spot spanwise tips. The additional mechanism induced by advection that we have revealed is worth testing in other geometries (plane Poiseuille, boundary layer \ldots) while the edge vortices and large scale flows structure and intensity should be studied more quantitatively to fully validate the scenario we have proposed here.\\
\paragraph*{}
We would like to thank P. Manneville and Y. Duguet for stimulating and helpful discussions and for their comment on this manuscript. Y. Zhu is also acknowledged for his contribution. This project was supported by ANR Jeunes chercheurs/Jeunes chercheuses QANCOUET.
\end{document}